\documentclass[10pt,conference]{IEEEtran}
\IEEEoverridecommandlockouts
\usepackage{cite}
\usepackage{amsmath,amssymb,amsfonts}
\usepackage{algorithmic}
\usepackage{graphicx}
\usepackage{textcomp}
\usepackage{xcolor}
\usepackage{multirow}
\usepackage{listings}
\usepackage{booktabs}
\usepackage{soul}
\usepackage[framemethod=tikz]{mdframed}
\usepackage{tikz}

\def\BibTeX{{\rm B\kern-.05em{\sc i\kern-.025em b}\kern-.08em
    T\kern-.1667em\lower.7ex\hbox{E}\kern-.125emX}}

\makeatletter 
\newcommand{\linebreakand}{%
  \end{@IEEEauthorhalign}
  \hfill\mbox{}\par
  \mbox{}\hfill\begin{@IEEEauthorhalign}
}
\makeatother 

\definecolor{valcolor}{HTML}{067D17}
\definecolor{backcolor}{HTML}{FFFFFF}
\definecolor{keycolor}{HTML}{0033B3}
\definecolor{attrcolor}{HTML}{174AD4}
\definecolor{comcolor}{HTML}{8C8C8C}

\makeatletter
\lst@CCPutMacro
    \lst@ProcessOther{"3C}{\lst@ttfamily{\textcolor{keycolor}{<}}{\textcolor{keycolor}{<}}}
    \lst@ProcessOther{"3E}{\lst@ttfamily{\textcolor{keycolor}{/> }}{\textcolor{keycolor}{/>}}}
    \lst@ProcessOther{"2A}{\lst@ttfamily{\textcolor{keycolor}{>}}{\textcolor{keycolor}{>}}}
    \lst@ProcessOther{"26}{\lst@ttfamily{\textcolor{keycolor}{</ }}{\textcolor{keycolor}{</ }}}
    \lst@ProcessOther{"3D}{\lst@ttfamily{\textcolor{valcolor}{=}}{\textcolor{valcolor}{=}}}
    \lst@ProcessOther{"21}{\lst@ttfamily{\textcolor{comcolor}{<!-- }}{\textcolor{comcolor}{<!-- }}}
    \lst@ProcessOther{"2D}{\lst@ttfamily{\textcolor{comcolor}{-->}}{\textcolor{comcolor}{-->}}}
    \@empty\z@\@empty
\makeatother

\lstdefinelanguage{XML}{
    morestring=[b]",
    morekeywords={action, typing, scroll, log, gaze, psi, level},
    morecomment=[s]{!}{-},
    basicstyle=\linespread{1.15}\footnotesize\ttfamily, 
    backgroundcolor=\color{backcolor},
    keywordstyle=\color{keycolor},
    stringstyle=\color{valcolor},
    identifierstyle=\color{attrcolor},
    commentstyle=\color{comcolor},
}

\mdfdefinestyle{KeyFinding}{
    backgroundcolor=gray!20, 
    roundcorner=5pt,
}

\title{A Study on Developer Behaviors for Validating and Repairing LLM-Generated Code Using Eye Tracking and IDE Actions}
\author{\IEEEauthorblockN{Ningzhi Tang\textsuperscript{*}\thanks{*Both authors contributed equally to this research.}}
\IEEEauthorblockA{
\textit{University of Notre Dame}\\
Notre Dame, IN, USA \\
ntang@nd.edu}
\and
\IEEEauthorblockN{Meng Chen\textsuperscript{*}}
\IEEEauthorblockA{\textit{University of Notre Dame}\\
Notre Dame, IN, USA \\
mchen24@nd.edu}
\and
\IEEEauthorblockN{Zheng Ning}
\IEEEauthorblockA{\textit{University of Notre Dame}\\
Notre Dame, IN, USA \\
zning@nd.edu}
\and
\IEEEauthorblockN{Aakash Bansal}
\IEEEauthorblockA{\textit{University of Notre Dame}\\
Notre Dame, IN, USA \\
abansal1@nd.edu}
\and
\linebreakand
\IEEEauthorblockN{Yu Huang}
\IEEEauthorblockA{\textit{Vanderbilt University}\\
Nashville, TN, USA \\
yu.huang@vanderbilt.edu}
\and
\IEEEauthorblockN{Collin McMillan}
\IEEEauthorblockA{\textit{University of Notre Dame}\\
Notre Dame, IN, USA \\
cmc@nd.edu}
\and
\IEEEauthorblockN{Toby Jia-Jun Li}
\IEEEauthorblockA{\textit{University of Notre Dame}\\
Notre Dame, IN, USA \\
toby.j.li@nd.edu}
}

\IEEEaftertitletext{\vspace{-2em}}

\begin{document}

\maketitle

\begin{abstract}
The increasing use of large language model (LLM)-powered code generation tools, such as GitHub Copilot, is transforming software engineering practices. This paper investigates how developers validate and repair code generated by Copilot and examines the impact of code provenance awareness during these processes. We conducted a lab study with 28 participants, who were tasked with validating and repairing Copilot-generated code in three software projects. Participants were randomly divided into two groups: one informed about the provenance of LLM-generated code and the other not. We collected data on IDE interactions, eye-tracking, cognitive workload assessments, and conducted semi-structured interviews. Our results indicate that, without explicit information, developers often fail to identify the LLM origin of the code. Developers generally employ similar validation and repair strategies for LLM-generated code, but exhibit behaviors such as frequent switching between code and comments, different attentional focus, and a tendency to delete and rewrite code. Being aware of the code’s provenance led to improved performance, increased search efforts, more frequent Copilot usage, and higher cognitive workload. These findings enhance our understanding of how developers interact with LLM-generated code and carry implications for designing tools that facilitate effective human-LLM collaboration in software development.
\end{abstract}


\section{Introduction}
\label{sec:intro}


Recent advances in large language models (LLMs) have significantly impacted the programming process, becoming increasingly integrated into the programming workflow~\cite{camara2023assessment}. A notable example is GitHub Copilot\footnote{https://github.com/features/copilot/}, a code generation tool powered by OpenAI's GPT model~\cite{brown2020language}, which can generate lines or subroutines of code based on existing code or natural language comments~\cite{chen2021evaluating}.  
These code generation tools are primarily used to enhance productivity, offering benefits such as autocomplete, quicker task completion, and easier recall of syntax~\cite {liang2023understanding}. In a controlled experiment, the group using Copilot completed programming tasks 55.8\% faster than the group that did not use it~\cite{peng2023impact}.

While LLM-powered tools facilitate code generation, the quality of the generated code is not guaranteed. A recent study found that LLM-generated code commonly suffers from code quality issues, such as compilation and runtime errors, incorrect outputs, and problems related to code style and maintainability~\cite{liu2023refining}. Consequently, after using these tools to generate a section of code, developers must spend time evaluating its correctness, fixing potential bugs, and integrating the code into the existing codebase. Previous studies have shown that developers spend a significant portion of their time (37.99\% in total) thinking/verifying suggestions, debugging/testing, and editing written code when programming with Copilot~\cite{mozannar2022reading}. 

Error discovery and repair strategies are crucial for successful human-AI interaction, guiding the development of underlying AI models and user interfaces, as indicated in other contexts such as conversational agents~\cite{li_sovite:_2020}, qualitative coding~\cite{gebreegziabher_patat:_2023}, and natural language data queries~\cite{ning_empirical_2023}. For code generation, AI tools can generate different types of errors than humans~\cite{dakhel2023github}, and unlike humans, AI cannot articulate the rationale behind its decisions. Therefore, the focus of attention and strategies that developers follow may differ from those used when working with human-written code. Although studies have explored how developers interact with such AI tools to generate code~\cite{liang2023understanding, vaithilingam2022expectation, barke2022grounded}, and the quality and usability of the generated code~\cite{chen2021evaluating, al2022readable}, the process of validating and repairing them is still not well understood, warranting further investigation.

Furthermore, as LLM-generated code becomes increasingly integrated with existing codebases, it is important to consider its provenance (i.e., whether the code is LLM-generated or human-written). Previous research has revealed that awareness of code provenance impacts developers' behavior when interacting with code, even though they may not always be conscious of such biases~\cite{terrell2017gender, imtiaz2019investigating, huang2020biases}. In a study on code review, despite all patches being created by automatic program repair tools, some were labeled as human-written, while others were labeled as machine-generated. Participants showed different scanning patterns and attention distributions in response~\cite{bertram2020trustworthiness}. In the context of debugging, developers' awareness of code provenance may also impact their mental models of what is more likely to go wrong and how they may go wrong, affecting their strategies and performance. Understanding these differences can help developers better comprehend LLM-powered code generation tools, leading to the design of tools that establish appropriate trust and support better collaboration.


Prior studies on developers' interactions with LLM-powered code generation tools have largely depended on subjective analyses, such as post-analysis of audio/video recordings~\cite{vaithilingam2022expectation, barke2022grounded, mozannar2022reading}, surveys~\cite{liang2023understanding}, and interviews~\cite{vaithilingam2022expectation, barke2022grounded}. These methods may suffer from recall and observer biases~\cite{etelapelto1993metacognition, davis2023s}, and fall short of reliably capturing developers' cognitive processes~\cite{fry2012human}. In response to these limitations, recent research in software engineering has increasingly utilized eye tracking as an objective measure. Eye tracking offers biologically-based insights into cognitive processes through non-intrusive observation of developers' gaze patterns~\cite{rodeghero2015empirical, bertram2020trustworthiness}. Eye-tracking data, specifically developer gaze patterns, can be used to infer their visual attention strategies~\cite{bednarik2006eye, aschwanden2006code} and may correlate with their self-reported cognitive workload~\cite{palinko2010estimating, zagermann2016measuring}. This provides an important indicator of their mental model when performing programming tasks~\cite{bednarik2012expertise, rodeghero2014improving}. Previous research also suggests that developer interactions within the Integrated Development Environment (IDE) can complement eye-tracking data to comprehensively understand their behaviors~\cite{hejmady2012visual, kazemitabaar2023novices}. Therefore, we employ a mixed-method approach, combining quantitative (i.e., eye tracking and IDE tracking) and qualitative (i.e., surveys and interviews) methods to understand developer behaviors in our study.

This paper addresses the gap in understanding how developers validate and repair LLM-generated code, specifically focusing on chunks of code (about tens of lines) produced by GitHub Copilot. We also examine the effects of informing developers about the code provenance on their validation and repair strategies. We conducted lab studies with 28 participants, tasking them with validating, repairing, and integrating Copilot-generated code into three representative software projects. Participants were allowed to interact with Copilot during the process. To compare developer behaviors under different awareness of the provenance of LLM-generated code, we randomly divided the participants into two groups: one was informed before the experiment that the code was generated by Copilot, while the other group was not informed.

\begin{figure}[htbp]
    \centerline{\includegraphics[width=0.5\textwidth]{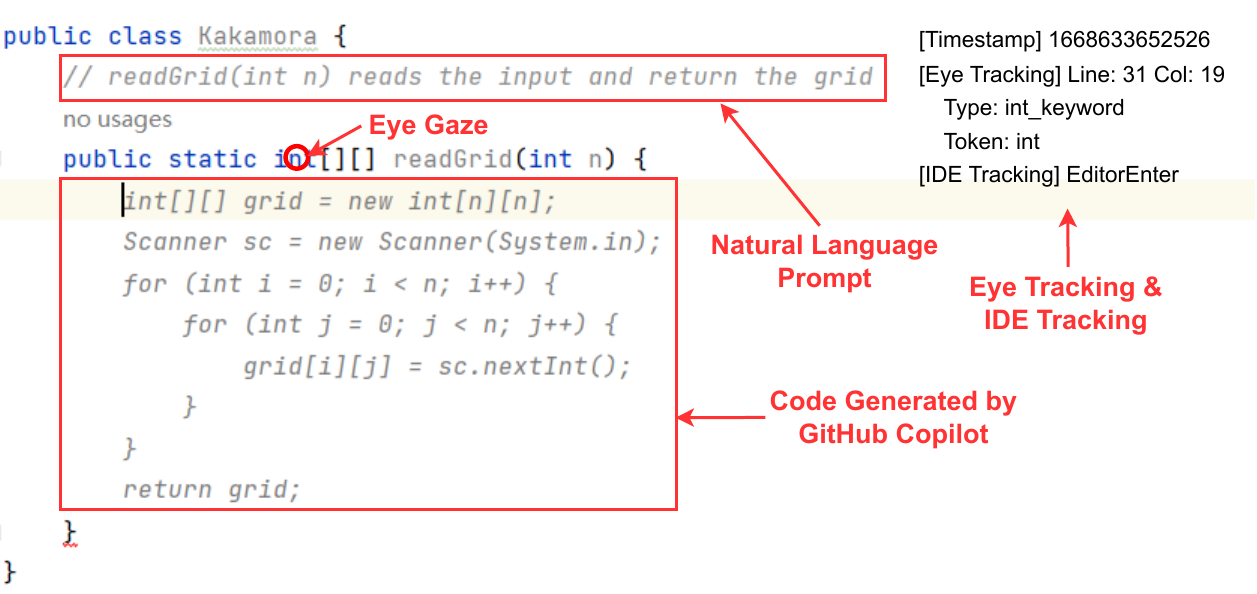}}
    \caption{A glance of how Copilot is invoked in our study setting, along with the collected IDE tracking and eye tracking data.}
    \label{fig:copilot_glance}
\end{figure}

During the validation and repair process of participants, we collected their IDE interaction and eye-tracking data using CodeGRITS~\cite{tang2024codegrits}, an IntelliJ platform plugin to collect developer behaviors\footnote{https://codegrits.github.io/CodeGRITS/}. After each subtask, we asked participants to report their cognitive workload using the NASA-TLX questionnaire~\cite{hart1986nasa}. After completing all three tasks, we conducted a semi-structured interview to discuss their mental models and strategies for validating and repairing codes. Figure~\ref{fig:copilot_glance} provides an overview of our data collection methods. We investigated the following research questions.
\begin{itemize}
    \item \textbf{RQ1.} What is developers’ perception of LLM-generated code compared to human-written code?
    \item \textbf{RQ2.} What are the differences in the strategies of developers to validate and repair LLM-generated code versus human-written code?
    \item \textbf{RQ3.} How does awareness of code provenance affect the validation and repair behaviors of developers?
\end{itemize}

We highlight the following findings of our study.
\begin{itemize}
    \item Developers often do not recognize the provenance of code unless explicitly informed. LLM-generated code typically features more comments, exhibits different error types compared to human-made errors, and demonstrates stronger syntax but weaker logic performance.
    \item While developers generally apply similar strategies for validating and repairing LLM-generated code as they do with human-written code, they also display distinct behaviors. These include frequent switching between code and comments, varied focus areas, and a tendency to delete and completely rewrite the code.
    \item Informing developers about the provenance of LLM-generated code positively influences their performance in validation and repair tasks. It leads to improved validation and repair outcomes, increased search efforts, and more frequent utilization of Copilot during these processes. However, this awareness also increases cognitive workload, as developers pay more attention to potential errors and integration challenges.
\end{itemize}

\section{Related Work}


\subsection{Empirical Study of LLM Code Generation Tools}

Recent advances in transformer-based~\cite{vaswani2017attention} LLMs marked a significant breakthrough in code generation. Several production-level code generation tools, such as GitHub Copilot, OpenAI GPT-4~\cite{achiam2023gpt}, Tabnine\footnote{https://www.tabnine.com/}, Amazon CodeWhisperer\footnote{https://aws.amazon.com/codewhisperer/}, and IntelliCode Compose~\cite{svyatkovskiy2020intellicode}, have been adopted by a growing number of developers~\cite{liang2023understanding}. Notably, Codex~\cite{chen2021evaluating}, a descendant of GPT-3 model~\cite{brown2020language} fine-tuned on 54 million public GitHub repositories, can solve 37.7\% of unseen Python code-writing tasks with a single sample in the HumanEval dataset~\cite{chen2021evaluating}, while GPT-4 has increased this to 67.0\%~\cite{achiam2023gpt}. Various metrics and datasets have been introduced to evaluate their performance in offline settings~\cite{chen2021evaluating, li2022competition, dakhel2023github}; however, more studies are needed to explore how developers use them and their impact on developers.

Previous studies explored how developers perceive and interact with these LLM code generation tools~\cite{ziegler2022productivity, vaithilingam2022expectation, wu2023ai, amoozadeh2024trust}. For example, Barke~\emph{et al.}~\cite{barke2022grounded} observed two modes of developer interactions: \textit{acceleration}, where developers use Copilot to code faster, and \textit{exploration}, where developers use it to explore what to do next. Liang~\emph{et al.}~\cite{liang2023understanding} conducted a large-scale survey with 410 developers and identified a set of usage characteristics and usability issues with AI programming assistants. Mozannar~\emph{et al.}~\cite{mozannar2022reading} proposed a taxonomy of common developer activities and retrospectively taxonomized 21 developers' coding sessions with Copilot. They found that developers spend more than half of their task time on Copilot-related activities and a large fraction of time validating and editing Copilot's suggestions. Our study focuses on how developers validate and repair Copilot-generated code, serving as a complementary perspective to the existing literature.

\subsection{Debugging Strategies for Software Developers}

Debugging is a complex process that involves various tasks, including code comprehension, error localization, and repair, and remains a long-standing topic in software engineering research~\cite{brooks1999towards, latoza2010developers, alaboudi2023constitutes}. Previous literature indicates that debugging involves iteratively comprehending the code and forming and testing hypotheses~\cite{araki1991general, gilmore1991models}. Lawrance~\emph{et al.}~\cite{lawrance2010programmers} modeled developers' debugging processes from an information foraging theory perspective and categorized six debugging modes: mapping, drill-down mapping, observe-the-failure, locate-the-fault, fix-the-fault, and verify. Liu~\emph{et al.}~\cite{liudebugging} observed professional developers debugging large-scale software systems and found that common debugging techniques generally fall into two modes: identify and fix the error. The usability of debugging tools has also been investigated. For instance, Beller~\emph{et al.}~\cite{beller2018dichotomy} observed limited use of IDE features and proposed improvements to the Eclipse debugger based on their findings.

However, LLM-generated code shows different traits from the human-written code. Previous studies have found that LLM-generated code is less compact~\cite{nguyen2022empirical}, lacking context~\cite{li2024enhancing}, and of lower quality than human-written code~\cite{imai2022github}. The process of developers debugging code that was often written by themselves can also be quite different from validating and repairing generated code.  Thus, our study is different in that we investigate how developers validate and repair LLM-generated code instead of human-written code. This contributes to a better understanding of the impact of LLM on software development and helps to develop better tools to alleviate new challenges.

\subsection{Eye Tracking in Software Engineering}

Eye tracking captures the visual attention of participants by recording their eye gaze data~\cite{just1980theory}. Visual attention triggers mental processes for comprehending and solving a given task, which provides information on the cognitive processes of the participants~\cite{rayner1978eye, may1990eye, zagermann2016measuring}. The recorded gazes are often computed as a time series of fixations (eye stabilization lasting more than 200 ms) and saccades (rapid movements between fixations within 50 ms) for analysis~\cite{sharafi2015eye}. Recently, researchers started using eye tracking to study the software development process, including code comprehension~\cite{busjahn2014influences}, review~\cite{bertram2020trustworthiness}, and debugging~\cite{hejmady2012visual}. For example, Rodeghero~\emph{et al.}~\cite{rodeghero2015empirical} studied the eye movement patterns of developers when summarizing code and found that these patterns were analogous to those observed in reading natural language. Sharafi~\emph{et al.}~\cite{sharafi2015eye} recommended a set of eye-tracking metrics for software engineering. Obaidellah~\emph{et al.}~\cite{obaidellah2018survey} conducted a survey on the use of eye tracking in software development in 2018. Eye tracking can also typically be integrated with tracking developers' interactions within the IDE (e.g., mouse clicks, file navigation, running the class, etc.) to understand how they develop software~\cite{ali2015empirical, hejmady2012visual, maan2020representational}. In this study, we use CodeGRITS~\cite{tang2024codegrits} to simultaneously track developers' IDE interactions and eye gazes for data analysis.

Previous studies also used developer behavioral data to explore differences in software development behavior under different code provenance awareness, such as gender~\cite{terrell2017gender, imtiaz2019investigating} or human/machine distinctions~\cite{bertram2020trustworthiness}. For example, Huang~\emph{et al.}~\cite{huang2020biases} combined medical imaging and eye tracking data to study biases and differences associated with the gender of humans and machines in code review. They found that participants were more likely to accept pull requests from women and less likely to accept those from machines. Our research extends these studies by using IDE tracking and eye tracking to explore the impact of code provenance on validating and repairing LLM-generated code.

\section{Study Design}

We conducted a lab study with 28 participants to investigate how developers validate and repair code generated by Copilot. 
We used GitHub Copilot as a representative LLM-based code generation tool because it is specifically designed for coding, is widely used by developers~\cite{liang2023understanding}, and has previously been studied in other works~\cite{vaithilingam2022expectation, mozannar2022reading}.
In our study, each participant completed three programming tasks (details in Section~\ref{sec:tasks}). For each task, we created a project containing several declared but unimplemented subroutines (e.g., classes/methods). We then developed prompts that described the intended functionality of each subroutine and used Copilot to generate code for them. We deliberately constructed prompts for which Copilot would generate code containing representative errors of different types, as described in Section~\ref{sec:tasks}.

To ensure consistency between participants, we presented each participant with the same code, prompts, and Copilot-generated code. We then asked the participants to validate and repair the generated code with the assistance of Copilot. The participants were randomly assigned to two conditions. In one condition, the participants were informed that the code was generated using LLMs. Participants in the other condition were not informed about the provenance of the code.

The details of our study protocol and study design can be found in Section~\ref{sec:protocol} and Section~\ref{sec:participants}.

\begin{figure}[htbp]
    \centerline{\includegraphics[width=0.45\textwidth]{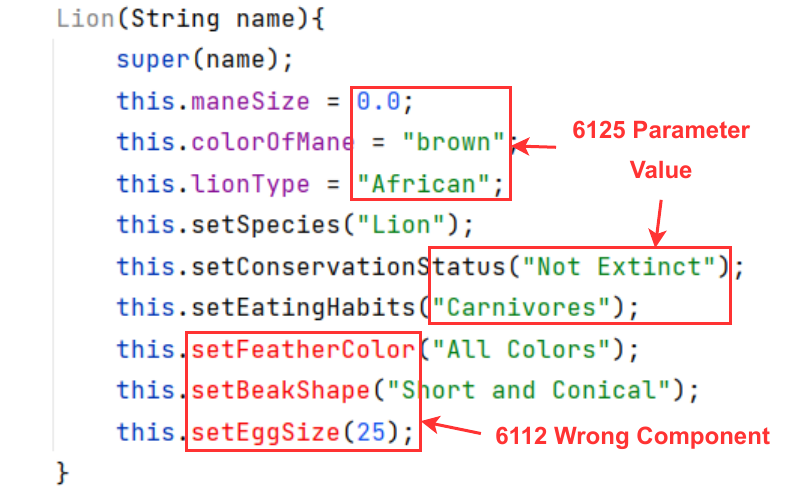}}
    \caption{An example of errors in LLM-generated code in task ZooSystem: 6112 (Wrong Component) and 6125 (Parameter Value).}
    \label{fig:example_bug}
\end{figure}

\subsection{Programming Tasks}
\label{sec:tasks}

We used three Java programming tasks in different software engineering scenarios: algorithm design, graphical user interface (GUI) design, and object-oriented programming (OOP).

\begin{itemize}
    \item \textbf{Task 1.} \textit{Kakamora (algorithm design)}: A LeetCode-like task that utilizes dynamic programming to find a path in a square array with the minimum sum of numbers. The entire program was written from scratch by Copilot. This task was adapted from an assignment in an undergraduate computer science course.
    \item \textbf{Task 2.} \textit{Calculator (GUI)}: A calculator app written using the Java GUI programming API. It consists of a front-end that includes a text interface and operational buttons, as well as back-end logic. The GUI layout code and the listeners for the front-end buttons were generated by Copilot.
    \item \textbf{Task 3.} \textit{ZooSystem (OOP)}: A zoo management system that includes various animal classes with inheritance relationships (e.g., \texttt{Animal}-\texttt{Mammal}-\texttt{Lion}). Several management functions, such as add, delete, search, and display animals, were generated by Copilot. This task was adapted from an assignment in an undergraduate computer science course.
\end{itemize}

We categorized these errors using the taxonomy presented by Beizer~\cite{beizer2003software}.
The error statistics for the three tasks are presented in Table~\ref{tab:bug_taxonomy}. Each subtask refers to a subroutine generated by Copilot that contains bugs, which may include multiple types of bugs (e.g., subtasks 1.2 and 3.2).

\begin{table}[htbp]
\centering
\caption{Bug types represented in programming tasks based on~\cite{beizer2003software}.}
\begin{tabular}{c|c|c|c}
    \toprule
    Task & Subtask & Bug Index & Bug Category \\
    \midrule \midrule
    \multirow{4}{*}{Kakaroma} & 1.1 & 231x & Missing Case \\
     & \multirow{3}{*}{1.2} & 3226.4 & String Manipulation-Insertion \\
     & & 3126 & Illogic Predicates \\
     & & 231x & Missing Case \\
    \midrule
    \multirow{2}{*}{Calculator} & 2.1 & 6125 & Parameter Value \\
     & 2.2 & 614x & Initialization State \\
    \midrule
    \multirow{6}{*}{ZooSystem} & 3.1 & 6125 & Parameter Value \\
     & \multirow{2}{*}{3.2} & 6112 & Wrong Component \\
     & & 6125 & Parameter Value \\
     & 3.3 & 4164 & Should be Dynamic Resource \\
     & 3.4 & 6112 & Wrong Component \\
     & 3.5 & 413x & Initial, Default Values\\
    \bottomrule
\end{tabular}
\label{tab:bug_taxonomy}
\end{table}

For example, Fig.~\ref{fig:example_bug} shows bugs in the code generated by Copilot for subtask 3.2, with the prompt ``create a constructor for the Lion class that only takes the name as input.'' The code has both bug type 6125 ``Parameter Value'' (e.g., the default conservation status should be \texttt{Vulnerable}), and bug type 6112 ``Wrong Component'' (\texttt{Lion} do not have feather; instead, it should be \texttt{setFurColor}, which extends from its superclass \texttt{Mammal}).

\subsection{Participants}
\label{sec:participants}

We recruited 28 participants (17 male, 11 female) from the local university community. Among them, 10 are graduate students and 14 are undergraduate students; 22 major in computer science and engineering, while two major in mathematics. Three participants had no prior programming experience in Java, but they had six, seven, and 12 years of experience in other programming languages, respectively. 12 had just completed an introductory course on Java programming, 12 had one to three years of Java programming experience, and one had over three years. However, regarding their general programming experience, participants on average had 5.5 years, and only three of them had less than three years of experience. Eight of them had used Copilot in the past before our study. The participants received a \$30 Amazon Gift Card as compensation for their time.

\subsection{Study Settings}

We conducted the study in person in a usability lab, using a Windows 10 computer equipped with a 27-inch monitor and a Tobii Pro Fusion eye tracker\footnote{https://www.tobii.com/products/eye-trackers/screen-based/tobii-pro-fusion}. The sampling frequency of the eye tracker was 60 Hz.


We randomly assigned the participants into two groups, each consisting of 14 participants, with different information about the code provenance. The \textbf{Informed} group was explicitly told that the code was generated by Copilot, while the \textbf{Non-Informed} group received no such indication.

During the study, participants used IntelliJ IDEA 2022.1.4, running the Copilot plugin and the data collector plugin (Section~\ref{sec:data_collection_setup}). We set the font size in the IDE to 20 points. To mitigate the influence of light intensity on eye tracking, all study sessions were held in the same room with all doors and windows closed and the same ceiling light on.

\subsection{Protocol}
\label{sec:protocol}

Each study session lasted approximately two hours. After the participant signed the consent form and completed a pre-study demographic questionnaire, we briefed the participant about the overall objective and process of the study, followed by a 10-minute tutorial on the use of IntelliJ IDEA and Copilot. We also gave a short instruction on interacting with the eye tracker (e.g., refraining from major head movements). 

For each task, the participant first read an instructional document that described the background of the tasks, followed by calibrating the eye tracker. The participant had then 20 minutes to perform each task. To narrow down the debugging strategy for specific errors, participants were asked to announce when they thought they had completed a subtask. Participants were not allowed to browse the Internet during the task. 


After completing each task, we asked participants to complete a NASA Task Load Index (NASA-TLX) questionnaire~\cite{hart1986nasa} to self-report their cognitive workload. NASA-TLX is a widely-used subjective assessment tool that measures the user's perceived workload when performing a task. It includes six dimensions: Mental Demand, Physical Demand, Temporal Demand, Performance, Effort, and Frustration. We used it to complement the data obtained from eye tracking.

At the end of the study session, we conducted a 20-minute semi-structured interview with the participants to discuss the behaviors we observed. The interview questions included their evaluations of the quality of LLM-generated code, their perceived differences between LLM-generated and human-written code, strategies they used to validate and repair bugs in LLM-generated code, their use of IDE features and Copilot, and their changes in mental models and strategies under different code provenances. We also asked participants from the Non-Informed group whether they realized that the code was generated by LLM during the study.

\subsection{Data Collection Setup}
\label{sec:data_collection_setup}

To support the analysis of developer behaviors, we used CodeGRITS~\cite{tang2024codegrits}, a plug-in for IntelliJ IDEA that enabled IDE tracking and eye tracking. In addition, we recorded the screen during study sessions with timestamps for subsequent analysis. The details of the collected data are summarized in the following sections.

\subsubsection{IDE Tracking}

CodeGRITS collects the following information via IDE tracking. All behaviors are tracked with their location (path, line, column) and timestamp for further analysis and calibration. They are organized in XML format, with an example shown in Listing~\ref{lst:ide_tracking}.

\textit{\textbf{IDE Features}} CodeGRITS recorded the use of various IDE features during programming. These include utilizing the clipboard, executing the program, employing a debugger (e.g., setting a breakpoint), switching among files, and navigating the code (e.g., \texttt{Find}, \texttt{GotoDeclaration}, and \texttt{FindUsages}). The usage log of IDE features also included actions related to the interaction with Copilot, such as \texttt{copilot.applyInlays}, which represents accepting the code generated by Copilot.

\textit{\textbf{Keyboard and Mouse Events}} CodeGRITS collected all keyboard events, including typing and navigating by keystrokes. It also detected keyboard events with special functions such as \texttt{Enter}, \texttt{Backspace}, or \texttt{Escape}. Additionally, the plugin recorded mouse scrolling behaviors, both vertical and horizontal.

\textit{\textbf{File Logging}} To enable the recovery of any code file at any time point during the tasks, we use CodeGRITS to record the entire content of the current code file whenever the developer opens, closes, or makes edits. Additionally, CodeGRITS records the console output whenever the developer executes the code.

\begin{lstlisting}[language=XML, caption={An example of IDE tracking data, including IDE features (top), keyboard events (middle), and file logging (bottom).}, label=lst:ide_tracking]

  !  IDE Features  -
<action id="ReformatCode" path="/src/Reptile.java"
        timestamp="1662928377436" >
<action id="copilot.applyInlays" path="/src/Lion.java" 
        timestamp="1662928415454" >
<action id="GotoDeclaration" path="/src/Lion.java"
        timestamp="1662928462655" >
<action id="RunClass" path="/src/ZooSystem.java"
        timestamp="1662928498642" >
  !  Keyboard Events  -
<action id="EditorPaste" path="/src/Lion.java" 
        timestamp="1662928472075" >
<typing column="4" length="13" line="8"
        path="/src/Reptile.java" string="System.out"
        timestamp="1662928450732" >
<scroll type="vertical" path="/src/ZooSystem.java" 
        offset="693" timestamp="1662928825827" >
  !  File Logging  -
<log id="fileLog" timestamp="1662928820593"
     info="fileOpened" path="/src/ZooSystem.java" >
<log id="fileLog" timestamp="1662928828525"
     info="fileChanged" path="/src/Lion.java" >
<log id="outputLog" timestamp="1662928831632" >
\end{lstlisting}

\subsubsection{Eye Tracking}
\label{sec:eye_tracking}
CodeGRITS collects raw gaze data and maps them to semantic source code entities, such as tokens and nodes in abstract syntax trees (ASTs). The sampling frequency was 60 Hz. A sample of gaze data and the semantic source code entities is shown in Listing~\ref{lst:eye_tracking}.

\textit{\textbf{Raw Gaze Data}} Each sample in the raw eye tracker data includes ($x, y$) relative coordinates on the screen, as well as the diameters of the pupils and their corresponding validity codes provided by the Tobii Pro SDK. We excluded invalid data from our analysis based on Tobii Pro's recommendations\footnote{https://developer.tobiipro.com/commonconcepts/validitycodes.html}. Each gaze point is recorded with a timestamp.

\textit{\textbf{Location, Token \& AST}} To extract semantic information from gazes, such as their corresponding code tokens, CodeGRITS calculated the relative location in the code editor, and then mapped it to specific locations (line, column) in the code file and their corresponding tokens. It then used IntelliJ's Program Structure Interface (PSI)\footnote{https://plugins.jetbrains.com/docs/intellij/psi.html} to determine the token, token type, and AST hierarchy associated with the gaze.

\begin{lstlisting}[language=XML, caption={An example of gaze data collected from the eye tracker, with the computed location, token, and AST hierarchy.}, label=lst:eye_tracking]

  !  Eye Tracking  -
<gaze x="545" y="306" path="/src/ZooSystem.java" 
      line="26" column="12" gaze_validity="1.0" 
      pupil_diameter="3.277" pupil_validity="1.0"
      start_timestamp="1669258026407" duration="286"
      end_timestamp="1669258026693" count="18"*
    <ast_type="IDENTIFIER" token="printSummaryView"*
        <level end="26:24" start="26:8" 
               tag="PsiIdentifier:printSummaryView" >
        <level end="26:24" start="26:8" 
               tag="PsiReferenceExpression" >
        <level end="26:31" start="26:8" 
               tag="PsiMethodCallExpression" >
        <level end="26:32" start="26:8" 
               tag="PsiExpressionStatement" >
        <level end="32:5" start="5:4"
               tag="PsiMethod:setupAnimals" >
        <level end="499:27" start="3:0"
               tag="PsiClass:ZooSystem" >
     &ast*
 &gaze*
\end{lstlisting}

\section{Research Methods}
In this section, we discuss the methods used to analyze the data collected from the study.

\subsection{Quantitative Analysis}

\subsubsection{Gaze Pattern Metrics}
\label{sec:gaze_pattern_metrics}

Fixation refers to stabilization of the eye at one location for a period of time, which is commonly used in eye-tracking studies and is often considered to be associated with visual attention and cognitive workload~\cite{dorr2010variability, sheridan2017chess}.
Following the practice recommended in previous research~\cite{sharafi2015eye}, we identified fixations by extracting gazes on the same token with durations longer than 200 ms, and we considered the transitions between two fixations within 50 ms as saccades. We computed the following metrics using the definitions provided in~\cite{sharafi2015eye} to analyze the cognitive processes of the participants.

\begin{itemize}
    \item \textit{Fixation Time}: Total duration of all fixations in seconds.
    \item \textit{Fixation Count}: Total number of fixations.
    \item \textit{Average Fixation Duration}: Average duration of each fixation in seconds.
    \item \textit{Saccade Time}: Total duration of all saccades in seconds.
    \item \textit{Saccade Count}: Total number of saccades.
    \item \textit{Average Saccade Length}: Average length of each saccade in pixels.
    \item \textit{Saccade Fixation Ratio}: Ratio of saccade time to fixation time, indicating the balance between exploration and exploitation during programming.
\end{itemize}

\subsubsection{Developer Behavior Categorization}
\label{sec:behavior_categorization}

\begin{table*}[htbp]
\centering
\caption{Categorization of developer behavior emerged from IDE and eye tracking data.}
\begin{tabular}{c|c|c}
    \toprule
    Index & Behavior & Tracking Data \\
    \midrule \midrule
    1 & Reading Document & Consecutive fixations on the instructional document \\
    2 & Reading Code & Consecutive fixations on the code \\
    3 & Reading Comment & Consecutive fixations on the comment \\
    \midrule
    4 & Switching Files & Opening, closing, or changing the selection of a file \\
    5 & Scrolling & Scrolling a file via mouse wheel, arrow keys, or touchpad gestures \\
    6 & Tracing Code & Searching tokens, finding usages or going to declarations \\
    7 & Running for Output & Running the class to obtain execution output \\
    8 & Employing Debugger & Utilizing debugger and its corresponding features (e.g., toggling breakpoints) \\
    9 & Invoking Copilot & Accepting, rejecting, or browsing code generated from Copilot \\
    10 & Utilizing Clipboard & Copying, cutting, or pasting contents \\
    11 & Keystrokes Typing & Typing characters using keystrokes \\
    \bottomrule
\end{tabular}
\label{tab:behavior_taxonomy}
\end{table*}

The data collector tracked rich low-level data within the IDE (Listing~\ref{lst:ide_tracking}). Among these, many can be aggregated for analysis. For example, \texttt{Run}, \texttt{RunClass}, and \texttt{Rerun} all indicate that the user ran the code to view output information. \texttt{Debug}, \texttt{StepOver}, and \texttt{ToggleLineBreakpoint} all represent behaviors related to using the debugger; whereas \texttt{ToggleInsertState} or \texttt{EditorLeftWithSelection} have no clear significance for understanding developer behavior. Therefore, we filtered and aggregated these low-level tracking data to achieve the high-level categorization shown in Table~\ref{tab:behavior_taxonomy}.

For eye tracking data, based on the fixations calculated in Section~\ref{sec:gaze_pattern_metrics}, we classified the tokens that the developer looks at into three types: code, comment, and document. This classification allowed us to identify three types of reading behaviors (Behavior 1-3 shown in Table~\ref{tab:behavior_taxonomy}).
For IDE interaction data, we first counted the number of occurrences for all the original data's ``id''. An author categorized them according to the similarity of their underlying activity types in initial high-level behaviors and discarded data that were not meaningful for understanding developer behavior. Subsequently, another author reviewed and, if necessary, revised the categorizations and definitions of behaviors. The disagreement cases were discussed to reconcile differences and establish a cohesive categorization (Behavior 4-11 shown in Table~\ref{tab:behavior_taxonomy}).

\subsection{Qualitative Analysis}



For the qualitative analysis of interview transcripts, we followed established open-coding procedures~\cite{brod2009qualitative, lazar2017research}. Initially, two authors independently performed qualitative coding on the transcripts. Subsequently, they discussed their findings to achieve agreement and formed a consolidated codebook. Using this codebook, we conducted a thematic analysis to identify emerging themes from the interviews. These themes were refined and developed into study findings.

\section{Study Result}
\label{sec:study_result}

In this section, we report the results for our three RQs. Except for Fig.~\ref{fig:behavior_patterns}, all calculations of frequency or time were averaged across each task for all participants.

\subsection{RQ1. Developers' Perceptions of LLM-Generated Code}
\label{sec:developer_perception}

We asked the fourteen participants from the Non-Informed group at the start of the interview, ``\textit{During the tasks, did you realize that the code you just validated was actually generated by LLM?}''
Almost 80\% (11/14) of them reported they did not realize it. This indicates that if not clearly informed, developers may not consider the provenance of the code, or they may implicitly assume that the code was written by a human. However, different perceptions of the code provenance impact validation and repair strategies, workload, and performance, which is further discussed in Section~\ref{sec:effect_provenance}.


Many participants (11 out of 28) indicated that the code had a good coding style and readability, even better than human-written code. For example, P21 reported, ``\textit{[LLM-generated code] does follow human formatting guidelines; variable names and everything were verbose and easy to use.}'' They stated that the LLM-generated code is clean and professional, because ``\textit{it writes better variable names or method names. In practice, many programmers tend to use very simple names.}'' (P25) Thus, ``\textit{it provides a pretty good starting point to validate.}'' (P10) P14, P15, P17, P21, P23, and P25 highlighted that LLM-generated code shows better understandability as it contains more detailed comments compared to human-written code.

However, participants reported that LLM made some mistakes that human developers would not. ``\textit{Sometimes I do not see why anyone would ever think that this was the right way to write this code.}'' (P24) For example, in Task 1, an algorithm design problem, Copilot used a series of if statements to hard-code the test samples. P19, P21, P24, and P26 all expressed confusion: ``\textit{When I saw the hardcoded `1', `2', and `3' in the path reconstruction, I was really confused and thought there would be some special meanings. But it turns out that AI just hard-coded the example output.}'' (P21). Regarding the bugs in Figure~\ref{fig:example_bug}, participants described them as ``\textit{ridiculous}'' and ``\textit{lacking common sense}'': ``\textit{Default values are given in the instructions, but the constructor has some errors. A lion cannot have a beak, feathers, or lay eggs.}'' (P16). 

Additionally, P10, P14, P26, and P28 also reported that the LLM-generated code often contains many repetitions. ``\textit{They put out the exact same code over and over again for every button, which is difficult to read through.}'' (P28). P26 expressed concerns about it: ``\textit{If I send this to my manager, they will reject my code; these repetitive for-loops are horrible.}'' Additionally, the LLM-generated code ``\textit{focuses too much on the prompts but loses the context of the code,}'' (P14) and ``\textit{may contain undefined variables,}'' (P10)

P6, P16, P25, and P27 also mentioned that errors in LLM-generated code and human-written code tend to manifest in different areas. ``\textit{Humans are more error-prone than Copilot when it comes to details; for the logic [of code], I think Copilot is more error-prone}''. (P25) This also affects their validation and repair approaches, which is discussed more in Section~\ref{sec:debugging_strategies}.

P14, P25, and P26 noted that Copilot lacks multimodal capabilities. For the GUI task in Task 2, a human developer would find ``\textit{easy,}'' but ``\textit{Copilot is not multimodal; it doesn't understand the layout of the calculator frontend. The text description lacks detail to accurately describe the layout of each button, which is probably why it just generates the buttons randomly.}'' (P25)

\begin{mdframed}[style=KeyFinding]
\textbf{Key findings:} If developers are not informed, they might not recognize that the code is generated by an LLM. LLM-generated code performs well in terms of coding style and readability. They also often include more comments. However, it tends to make unique mistakes that are uncommon for human developers. Such errors include repetitive structures, a lack of contextual understanding, and generally better performance in syntax than in logic. Additionally, LLMs' lack of multimodal capabilities, such as understanding visual layouts in front-end tasks, further differentiates the types of errors they are prone to make.
\end{mdframed}

\subsection{RQ2. Strategies to Validate and Repair Code}
\label{sec:debugging_strategies}

\begin{figure}[htbp]
    \centering
    \includegraphics[width=0.5\textwidth]{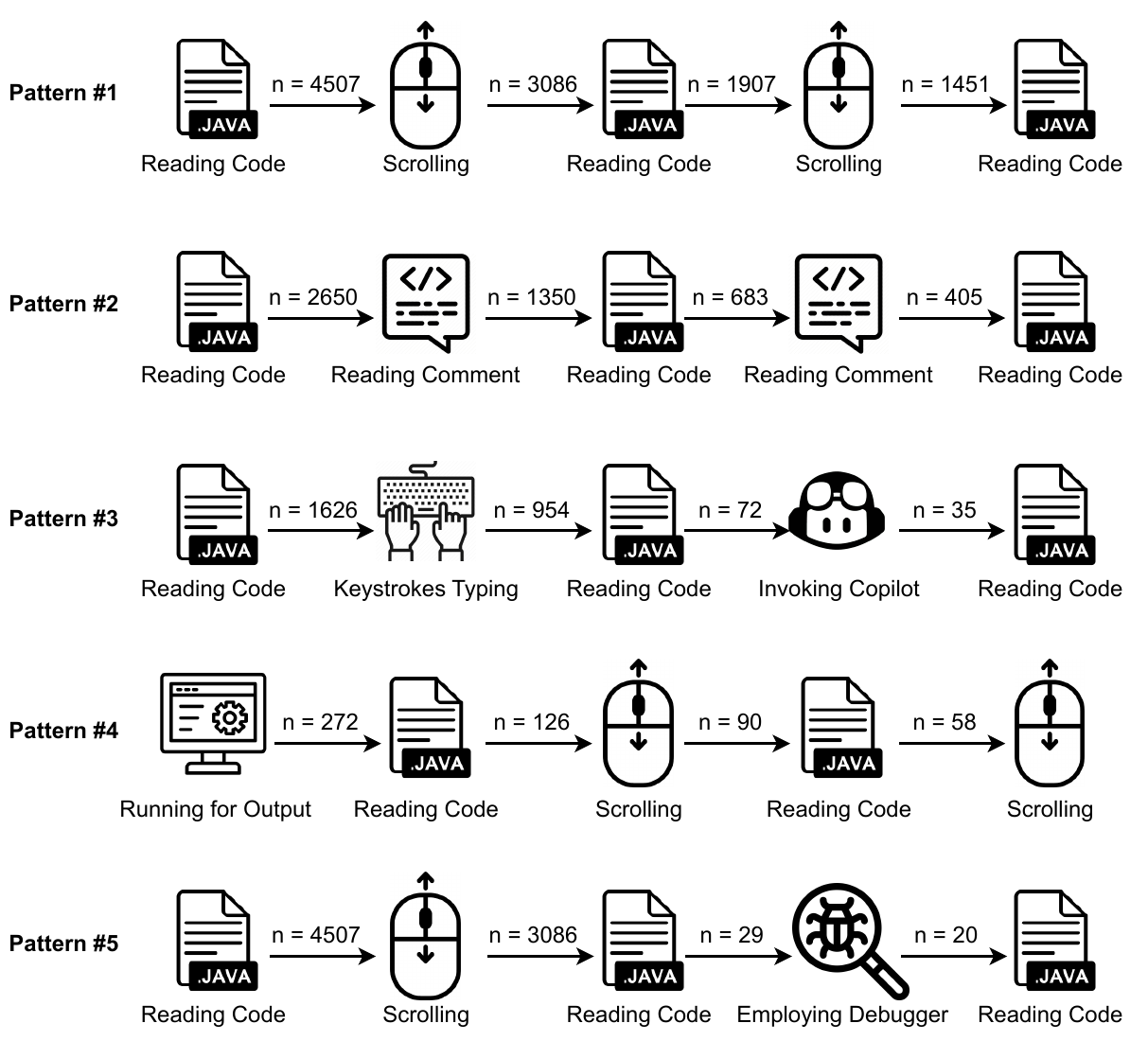}
    \caption{Common patterns of transitions among distinct behaviors across all participants. Patterns \#1 and \#2 are the top two most frequent patterns across all sequences. Patterns \#3, \#4, and \#5 are the most frequent patterns that involve Invoking Copilot, Running for Output, and Employing Debugger, respectively.}
    \label{fig:behavior_patterns}
\end{figure}

Based on the categorization of developer behaviors in Section~\ref{sec:behavior_categorization}, we calculated the frequency of transition sequences of behaviors and selected some high-frequency patterns in Fig.~\ref{fig:behavior_patterns}. Combining the findings of the interviews with tracked data, we identified the following validation and repair strategies used by the participants.

\subsubsection{General Strategies for Validation and Repair}

In general, participants reported in the interview that errors in LLM-generated code were found to be comparable to those made by humans, leading them to employ common validation and repair strategies as they normally would. The strategies they described are a mix of the following four aspects:
\begin{itemize}
    \item \textit{\textbf{Context Comprehension.}} Many participants (11/28) reported that they would first go through all the code to understand its overall functionality, then search for the cause of bugs. For example, P21 stated ``\textit{I have to read the initial code and try to understand what it does, and then understand why its way of doing things is wrong.}'' P20 said ``\textit{First I read through the code to conceptualize them in my mind.}'' This strategy aligns with previous theories about debugging, suggesting that an initial comprehension of the code helps form the context in the developer's mind, which serves as the basis for further bug locating and repairing~\cite{gilmore1991models, ko2006exploratory}.
    \item \textit{\textbf{Step-by-Step Validation.}} P24 and P28 reported that they would check the code's control flow step-by-step in mind to locate bugs. P24 states, ``\textit{I typically stare at the code for a good amount of time and try to understand what the variables are, like what's being sent where and their types. I just try to narrow down the lines as to where it might be going wrong so that I can find the lines that are causing the problem.}'' They also use some tools for tracing code in the IDEs to help navigate and understand the process. P17 states, ``\textit{I like to use `GotoDeclaration' and `FindUsages' to jump between sections}'' The data from IDE tracking also supports their reports. On average, participants used \texttt{GotoDeclaration} and \texttt{FindUsages} 1.81 times per task.
    \item \textit{\textbf{Running for Output.}} Participants often run the code to validate the presence and find the causes of bugs based on the output, as shown in Pattern \#4 of Fig.~\ref{fig:behavior_patterns}. On the one hand, this output is the final result required by the code, such as the testing of some samples in Task 1, or the visualization of the GUI in Task 2 -- ``\textit{Typically, I first run the program and see how it functions, especially for the GUI.}'' (P25). On the other hand, participants might manually insert some print statements to test the intermediate states of program execution, which is consistent with previous studies~\cite{araki1991general}. For example, P24 stated, ``\textit{I typically use if statements and print statements frequently to determine whether this condition is true at this time.}'' P28 also reported ``\textit{Copilot was very convenient to write for loops that print everything for me.}'' 
    \item \textit{\textbf{Employing Debugger.}} As shown in Pattern \#5 of Fig.~\ref{fig:behavior_patterns}, P7, P23, P25, and P26 also report using the default debugger in IDE to help with this process, i.e., toggling breakpoints and stepping through the code. However, inserting print statements and then running the entire code is much more common (7.483 $>$ 2.362, Student's t-test, p-value = 3.45e-5 $<$ 0.01); this might be due to their unfamiliarity with the IDE debugger~\cite{beller2018dichotomy}.
\end{itemize}

\subsubsection{Difference for LLM-Generated Code}
\label{sec:strategy_difference}

Although the validation and repair of LLM-generated code is reported to be similar to that of human-written code, we also observe some differences. First, as shown in Pattern \#2 of Fig.~\ref{fig:behavior_patterns}, developers tend to switch fixations consecutively between code and comment. Since the comments in the task projects are mainly used as prompts for Copilot to generate buggy code, we hypothesize that developers are trying to disambiguate the mismatch between the code and the prompt. For example, P14 states, ``\textit{There's a gap between the comment and the code to validate}''; P13 states, ``\textit{I need to match the expected output [code] to the input [prompt] provided to Copilot.}'' Otherwise, P3 also reported, ``\textit{Switching between instructions and code is annoying and challenging.}'' This phenomenon was also observed in other studies like~\cite{sarkar2022like} and~\cite{yen2023coladder}.

Second, participants also reported that LLM-generated code might offer a different perspective from human thoughts, potentially making validation and repair easier. P4 stated, ``\textit{Copilot has a completely different thinking angle; debugging my own code, I would fall into the same traps,}'' while P23 stated, ``\textit{Copilot can get things wrong as well, but it usually errs in a different way. From the two together, you can find the right solution.}''

Third, some LLM-generated code contains numerous erroneous statements that are difficult to fix by modifying only small parts. Participants tend to delete or comment out almost all of them and rewrite the code themselves. For example, P17 stated, ``\textit{With the layout, I just completely threw out the existing layout and kind of made one that worked myself.}'' P25 mentioned, ``\textit{Especially for the first task, I must say this is not a complex task, but I spent a lot of time figuring out the details [of the LLM-generated code]. [...] If I can do it next time, maybe I will just delete the entire code and write it myself.}'' 

Fourth, As stated in Section~\ref{sec:developer_perception}, participants think that LLM is more likely to make errors in the overall logic of the code, while humans are more likely to make mistakes in the details. For example, P25 states that ``\textit{I will focus more on the logic of the code if it is generated by LLM, probably because I don't think Copilot can generate really complicated logic}''; LLM rarely makes mistakes in generating similar, repetitive code blocks -- ``\textit{If you use LLM to generate similar code, it tends to be either all correct or all incorrect.}'' Therefore, ``\textit{If it is a similar bunch of code, I would trust the LLM; as for logic details, I will double-check the correctness of the LLM-generated code.}'' P27 also stated, ``\textit{[I] focus 100\% more on the structure. I suppose LLM would do better with syntax.}''

Lastly, another challenge of validating and repairing LLM-generated code is not knowing the reasoning behind LLM's decisions and not being able to communicate with the author. P26 stated, ``\textit{If I write the code myself, I generally know the logical thought process behind it. [...] I don't see how LLM is thinking about it.}'' P1 stated, ``\textit{I can ask a human coder about their ideas, but I don't know how Copilot generates its code.}''

\subsubsection{Role of Copilot in Validation and Repair}

In our study, we allowed participants to use Copilot in the validation and repair process. They reported that using Copilot itself was helpful. On average, participants used Copilot to generate more characters than through keystroke typing (362.84 $>$ 131.31, Student's t-test, p-value = 0.0007 $<$ 0.01). However, there is no statistically significant difference between the characters generated by Copilot and those from clipboard operations, i.e., copy-paste or cut-paste (362.84 $>$ 324.41, Student's t-test, p-value = 0.86 $>$ 0.10). 

Participants reported that Copilot facilitates the process of idea exploration, provides syntax suggestions while accelerating the speed of coding. For example, P7 stated, ``\textit{I change my approach. [...] First, I see if Copilot can give me some useful information, then I validate it.}''; P2 stated, ``\textit{If I don't know anything, I can just use Copilot to generate code; it gives me an idea of what to do next.}''; P24 commented, ``\textit{[AI is] very helpful when I don't know the Java syntax, [like] how to convert an ArrayList to an Array.}''; P4 mentioned, ``\textit{Copilot could replace Google searches and original documentation; you describe what you want and it understands what you mean.}'' These findings are also consistent with previous research~\cite{barke2022grounded, liang2023understanding}. 


Notably, we also observed that participants used Copilot to generate inline comments to help understand the code itself. P25 states, ``\textit{I think if that line of code has a bug, generating the comments directly from the code will help me figure it out.}''

\begin{mdframed}[style=KeyFinding]
\textbf{Key findings:} Developers generally apply similar validation and repair strategies to both LLM-generated and human-written code, which include context comprehension, step-by-step validation, running tests for output, and using debuggers. However, our study identified several new behaviors specific to handling LLM-generated code: developers frequently switch between code and comments, often completely delete and rewrite code, and exhibit a shifted focus from syntax to logic. Copilot was frequently used to enhance the process, particularly through the generation of inline comments that help clarify the code's functionality.
\end{mdframed}

\subsection{RQ3. Effects of Code Provenance Informing}
\label{sec:effect_provenance}

In this section, we investigate the impact of code provenance, i.e., whether the code was written by LLM or humans, on developers. We compare the differences between the Informed and Non-Informed groups based on our tracked data.

\subsubsection{Bug Fixing Success Rate}

\begin{figure}[htbp]
    \centering
    \includegraphics[width=0.5\textwidth]{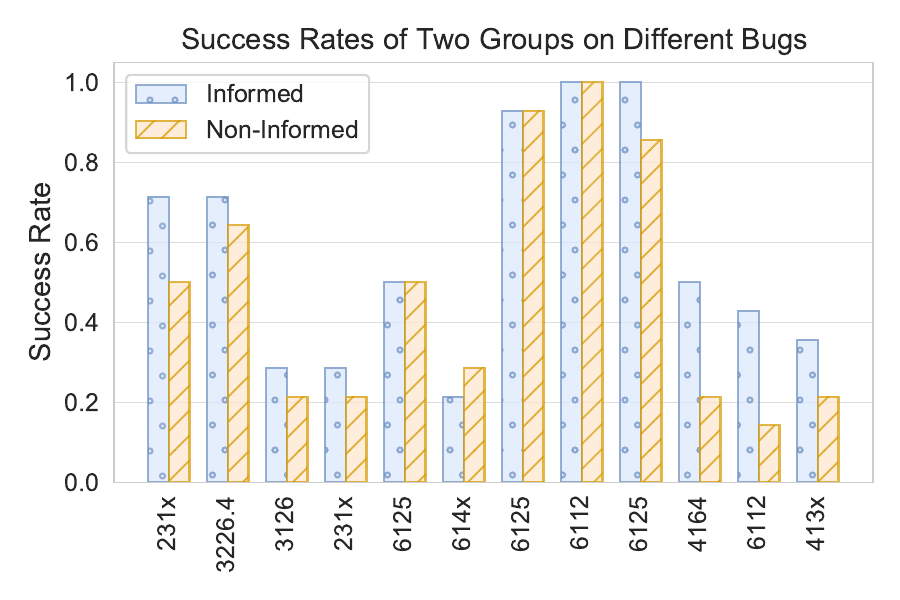}
    \caption{The success rates of bug validating and repair of the Informed group and the Non-Informed group.}
    \label{fig:success_rate}
\end{figure}

We first analyze the success rate of fixing the bugs in Table~\ref{tab:behavior_taxonomy} for each group. As shown in Fig.~\ref{fig:success_rate}, we observed that the Informed group performs better than the Non-Informed group (average success rate 0.577 $>$ 0.446, Student's t-test, p-value = 0.0417 $<$ 0.05). When participants were informed that the code was actually written by Copilot, compared to those who were not informed, they performed better on 8 bugs, the same on 3 bugs, and worse on 1 bug. Thus, the reasons behind such a different success rate warrant further exploration.

\subsubsection{Developer Behaviors}

As stated in Section~\ref{sec:strategy_difference}, participants showed a high frequency of fixation switching between code and comments. Thus, we first computed the ratio of time spent on code to comments to see if there were any differences between the two groups. These metrics can be inferred from eye-tracking data, i.e., Behaviors 1-3 in Table~\ref{tab:behavior_taxonomy}. The results suggest that the Informed group had a higher Time on Code/Time on Comment ratio, which is partially significant (0.233 $>$ 0.179, Student's t-test, p-value = 0.093 $<$ 0.10). The Pearson correlation analysis also showed that the higher ratio was positively correlated with the success rates across 12 bugs for each person (0.103 with a p-value of 0.066 $<$ 0.10). 

Additionally, the Informed group showed significantly less time reading the original document (48.6 $<$ 67.4, Student's t-test, p-value = 0.029 $<$ 0.05). Since the comments mainly consist of task content mentioned in the document and serve as prompts for Copilot to generate buggy code, these results indicate that developers focus more on the prompts given to the LLM when they know the code was generated by LLM. 

\begin{figure}[htbp]
    \centering
    \includegraphics[width=0.48\textwidth]{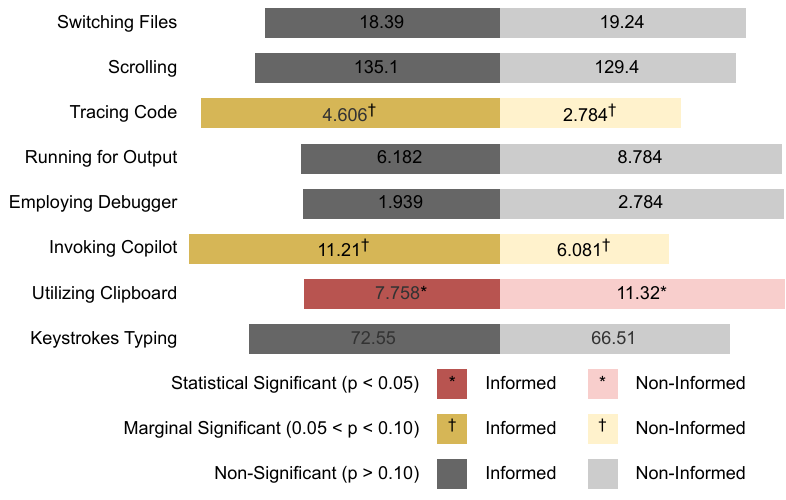}
    \caption{Comparison of the frequency of different types of developer behaviors between the Informed group (left) and the Non-Informed group (right).}
    \label{fig:ide_comparison}
\end{figure}

As shown in Fig.~\ref{fig:ide_comparison}, we calculated the frequency of developer behaviors, excluding reading (Behaviors 4-11 in Table~\ref{tab:behavior_taxonomy}), for both groups. From the (partially) statistically significant results, we make the following two observations:
\begin{itemize}
    \item If informed that the code was generated by LLM, participants use tracing code features (e.g., Find, GotoDeclaration, FindUsages) more frequently (4.606 $>$ 2.784, Student's t-test, p-value = 0.078 $<$ 0.10). Figure~\ref{fig:workload_comparison} in Section~\ref{sec:cognitive_workload} also shows that they have higher Saccade Time (2.303 $>$ 1.705, Student's t-test, p-value = 0.076 $<$ 0.10), which indicate greater search effort~\cite{alex2005eye}. This may be because they focus more on the high-level logic of the code than on the low-level details, as stated in Section~\ref{sec:strategy_difference}.
    \item If informed that the code is generated by LLM, participants use Copilot more (11.21 $>$ 6.081, Student's t-test, p-value = 0.093 $<$ 0.10) and the clipboard less (7.758 $<$ 11.32, Student's t-test, p-value = 0.017 $<$ 0.05). Note that most of the participants (20 out of 28) had not used Copilot prior to the study, but they were evenly distributed into two groups (10 in each), so the experimental results should not be affected by this factor. 
\end{itemize}


\subsubsection{Cognitive Workload}
\label{sec:cognitive_workload}

\begin{figure}[htbp]
    \centering
    \includegraphics[width=0.48\textwidth]{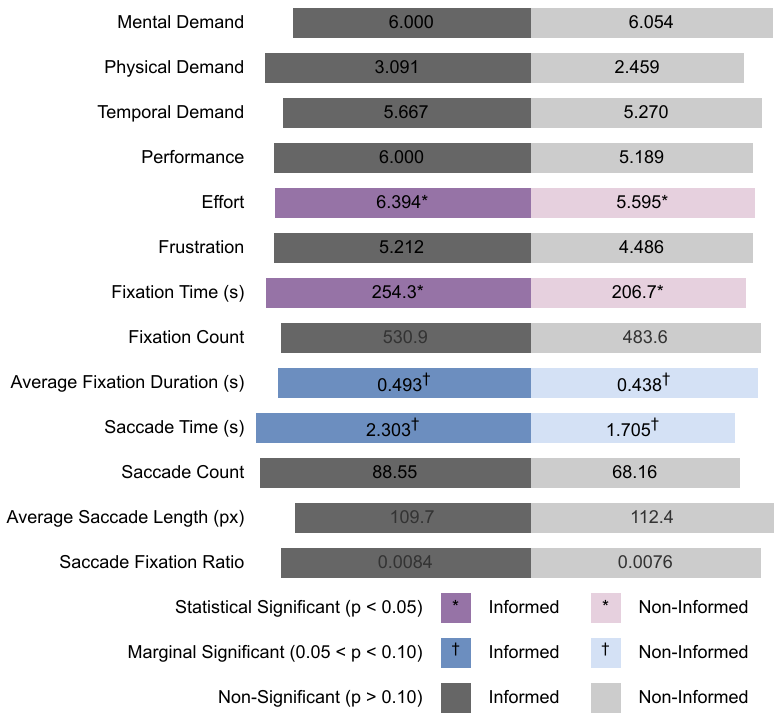}
    \caption{Comparison of self-reported workload measured by NASA-TLX and eye-tracking-related metrics between the Informed group (left) and Non-Informed group (right).}
    \label{fig:workload_comparison}
\end{figure}

We also calculated the workload metrics based on NASA-TLX~\cite{hart1986nasa} and the gaze pattern metrics described in Section~\ref{sec:gaze_pattern_metrics}. As shown in Fig.~\ref{fig:workload_comparison}, the Informed group had higher self-reported effort (6.394 $>$ 5.595, Student's t-test, p-value = 0.042 $<$ 0.05), fixation time (254.3 $>$ 206.7, Student's t-test, p-value = 0.050 = 0.05), and average fixation duration (0.493 $>$ 0.438, Student's t-test, p-value = 0.090 $<$ 0.10). Both fixation time and average fixation duration indicate greater visual attention and cognitive workload, as demonstrated in previous studies~\cite{dorr2010variability, sheridan2017chess}. The results suggest that developers who are informed that the code is generated by LLM tend to have a higher cognitive workload.

However, there is controversy among participants regarding their understanding of the impact of code provenance awareness on cognitive workload. For example, P28 stated, ``\textit{I think I would be more concerned [if I am aware of the provenance]. I would have to be more careful in figuring out what's going on. I don't trust LLM enough for an entire project.}'' Meanwhile, P25 states, ``\textit{I think the workload will be reduced if I know the code is generated by LLM because I only need to focus on some specific logic details instead of focusing on a similar bunch of code.}'' Since our study design is a between-subjects study, participants' guesses are based on their previous code validation and repair experience and may have recall biases~\cite{davis2023s}.

\begin{mdframed}[style=KeyFinding]
\textbf{Key findings:} When developers are informed that the code is generated by LLM, they perform better in validating and repairing the code. Subsequently, they tend to spend more time examining the prompts provided to Copilot compared to the code, and spend less time reading documents. When informed, developers make a greater search effort by using tracing code features more frequently and exhibit higher saccade times. They also used Copilot more frequently and used the clipboard less frequently. Finally, they experience a higher cognitive workload, as indicated by self-reported effort, fixation time, and average fixation duration.
\end{mdframed}

\section{Implications}

\textit{\textbf{Code provenance awareness.}} Developers may not realize that the code is generated by LLMs unless explicitly informed. Awareness of code provenance enhances their ability to effectively validate and repair code, likely due to a more focused approach to the generation prompts and a better understanding of potential issues to anticipate. This suggests the importance of labeling LLM-generated code, which could assist developers in better managing such code and inform the development of automated tools for recognizing and validating generation prompts. Additionally, for LLM-generated code, developing better tools to help developers understand and revise the underlying prompts could be beneficial.

\textit{\textbf{Improved logic in generated code.}} LLM-generated code typically excels in syntax but falls short in logic, prompting developers to focus differently during validation and repair tasks. Future tools could benefit from aiding developers in understanding and managing the logic of generated code. One potential approach is to design interfaces that allow developers greater control over the logic, while LLMs handle the syntax. Tools like CoLadder~\cite{yen2023coladder}, which supports hierarchical task decomposition of prompts before generating code, exemplify opportunities for advancement in this area.

\textit{\textbf{Multimodal capability.}} LLMs currently lack the capability to understand the multimodal context of programming tasks, such as visual layouts in UI development. This limitation underscores the need for AI models and interfaces that can effectively integrate multimodal information into code generation processes. Enhancing how generated code can be validated in conjunction with other modal information (e.g., audio, images, GUI layouts) is crucial for comprehensive development environments.

\textit{\textbf{Reducing switching costs.}} 
The frequent switching between code and comments/prompts observed during the validation and repair of LLM-generated code indicates a need for improved interfaces in development tools. Visualization tools that seamlessly connect prompts with generated code could reduce these switching costs, fostering a deeper and more efficient understanding for developers.

\section{Threats to Validity}

Our study faces several validity threats, one of which involves the selection of programming tasks. We chose three light-weight tasks in a popular programming language (i.e., Java), designed to represent typical types of programming tasks, and iteratively invoked Copilot to generate code with representative error types. Despite this, the tasks selected may not fully capture the complexity and diversity of real-world programming tasks that developers encounter.

Participant selection poses another threat to validity. Our participants consist of mostly undergraduate and graduate students in Computer Science, and their experience may not fully represent that of professional developers. However, the focus of our study was not on the impact of expertise, making the use of students a potentially acceptable option~\cite{kitchenham2002preliminary}. To enhance the ecological validity of our findings, a future longitudinal study in actual software development settings would be beneficial.


Furthermore, eye tracking often shows drift after prolonged use~\cite{sharafi2020practical}. To mitigate this effect, we performed a calibration before each task. To account for the validity of the conclusions, we used well-documented eye-tracking metrics and analyses~\cite{sharafi2015eye}. We also complemented the eye-tracking metrics with self-reported workload via NASA-TLX~\cite{hart1986nasa} as a supplement.

\section{Conclusion and Future Work}

We conducted a lab study with 28 participants to observe their behavior while validating and repairing LLM-generated code. Participants completed three Java programming tasks with erroneous code generated by GitHub Copilot. We gathered data through IDE tracking, eye-tracking, surveys, and semi-structured interviews.

Our findings indicate that, without explicit notification, developers often fail to recognize the provenance of the code, which can impact their performance, behaviors, and cognitive workload.  Although the LLM-generated code exhibits good readability and style, it introduces errors that differ from typical human-made errors. Although developers generally apply familiar validation and repair strategies, they also demonstrate unique behaviors in this context. These observations underscore the need for new tools and methods tailored to support developers in the LLM era.

In the future, we plan to systematically investigate the characteristics of LLM-generated code from a human perception perspective, complementing our behavior-centric approach. This effort will inform the development of more effective tools for developers working with LLM-generated code. Additionally, we plan to expand our research to include professional developers in long-term, real-world studies. We also intend to incorporate additional biometric sensors, such as heart rate monitors and fMRI, to gain a better understanding of the cognitive processes involved in software development.

\section*{Acknowledgment}
This research was supported in part by an AnalytiXIN Faculty Fellowship, an NVIDIA Academic Hardware Grant, a Google Cloud Research Credit Award, a Google Research Scholar Award, and NSF grants CCF-2211428 and CCF-2100035. Any opinions, findings, or recommendations expressed here are those of the authors and do not necessarily reflect the views of the sponsors. We thank Gelei Xu and Junwen An for useful discussion and valuable feedback on the project. We also thank Robert Wallace for his assistance in setting up the eye tracker.

\bibliographystyle{IEEEtran}
\bibliography{ref}

\end{document}